\documentclass{nature}
\usepackage{graphicx}
\usepackage{color}
\makeatletter
\let\saved@includegraphics\includegraphics
\AtBeginDocument{\let\includegraphics\saved@includegraphics}
\renewenvironment*{figure}{\@float{figure}}{\end@float}
\makeatother
\usepackage{amsmath} 
\usepackage{nccmath}
\usepackage{mathtools} 
\usepackage{multirow}
\newcommand{\angstrom}{\textup{\AA}}
\usepackage{booktabs}
\usepackage{array}
\usepackage{upgreek}
\usepackage{colortbl}
\usepackage{hhline}
\usepackage{boldline}
\usepackage{caption}
\usepackage{siunitx}
\usepackage{changes}
\usepackage[version=4]{mhchem}
\usepackage[hidelinks]{hyperref}
\usepackage{mathptmx}
\usepackage{cleveref}
\usepackage{xcolor}

\usepackage{caption} 
\usepackage{newfloat} 
\DeclareFloatingEnvironment[fileext=edf,placement=htbp,name=Extended Data Figure]{extendedfigure}

\title{Probing the atomic dynamics of ultrafast melting \\
       with femtosecond electron diffraction}

\author{M.\,Z.~Mo$^{1}$, M.\,B.~Maigler$^{2}$, T.~Held$^{3}$, 
        B.\,K.~Ofori-Okai$^1$, A.~Bergermann$^{1,4}$, Z.~Chen$^1$, 
        R.\,K.~Li$^{1}$, X.~Shen$^{1}$, K.~Sokolowski-Tinten$^5$, 
        R.~Redmer$^{4}$, X.\,J.~Wang$^{6,7}$, J.~Schein$^{2}$, 
        D.\,O.~Gericke$^{8}$, B.~Rethfeld$^{3}$, and S.\,H.~Glenzer$^1$}

\begin{document}
\maketitle

\begin{affiliations}
 \item SLAC National Accelerator Laboratory, Menlo Park, California, 94025, USA.
 \item Bundeswehr University Munich, 85577 Neubiberg, Germany
 \item Department of Physics and Research Center OPTIMAS, RPTU Kaiserslautern-Landau, 67663 Kaiserslautern, Germany
  \item Institut f\"ur Physik, Universit\"at Rostock, 18051 Rostock, Germany
  \item Faculty of Physics and Centre for Nanointegration Duisburg-Essen (CENDIE),
University of Duisburg-Essen, Lotharstrasse 1, 47048 Duisburg, Germany
  \item Faculty of Physics, University of Duisburg-Essen \& Research Center Chemical Sciences and Sustainability, 47048 Duisburg, Germany
 \item Department of Physics, TU Dortmund University, 44227 Dortmund, Germany
 \item Centre for Fusion, Space and Astrophysics, Department of Physics, University of Warwick, Coventry CV4 7AL, United Kingdom
\end{affiliations}

\newpage
\begin{abstract}
\textbf{Melting is an every-day phase transition that is determined by thermodynamic parameters like temperature and pressure.
In contrast, ultra-fast melting is governed by the microscopic response to a rapid energy input and, thus, can reveal the strength and dynamics of atomic bonds as well as the energy flow rate to the lattice.
Accurately describing these processes remains challenging and requires detailed insights into transient states encountered. 
Here, we present data from femtosecond electron diffraction measurements that capture the structural evolution of copper during the ultrafast solid-to-liquid phase transformations. 
At absorbed energy densities 2–4 times the melting threshold, melting begins at the surface slightly below the nominal melting point followed by rapid homogeneous melting throughout the volume.
Molecular dynamics simulations reproduce these observations and reveal a weak electron-lattice energy transfer rate for the given experimental conditions. 
Both simulations and experiments show no indications of rapid lattice collapse when its temperature surpasses proposed limits of superheating, providing
evidence that inherent dynamics limits the speed of disordering in ultrafast melting of metals.}
%

\end{abstract}

Rapid energy deposition into a material drives a cascade of kinetic and structural processes\cite{Sundaram2002, Shugaev2016}. 
The energy source, typically a short-pulse laser, usually couples almost entirely to the electrons driving them into a state far from equilibrium. 
For metals, the subsequent relaxation to Fermi-distributed electrons occurs on femtosecond time scales\cite{Hohlfeld2000,Mueller2013PRB, Chen2021}. 
Afterward, the material can often be modeled as a two-temperature system with hot electrons and a cold lattice\cite{Anisimov1974,Rethfeld2017}, but the electronic band occupation and nuclei positions may still not be in equilibrium\cite{Ndione2022SciRep,Stampfli1994}. 
The energy transfer from the electrons to the lattice is the slowest kinetic process and finally heats the lattice via electron-phonon coupling. 
Once the lattice reaches its melting temperature, one expects a phase transition, but 
nucleation kinetics competes with further lattice heating, and the loss of lattice structure may be considerably delayed\cite{Rethfeld2002melt,Ivanov03PRL,Siwick2003,Mo2018,Assefa2020,Antonowicz2024,White2025}. 

Understanding the chain of processes that leads to ultrafast melting of metals and, in particular, the time scales for all relaxation stages is crucial for a wide range of applications including laser micromachining\cite{Gattass2008}, studies of planetary interiors\cite{Guillot72,Pozzo2012}, 
inertial fusion experiments\cite{Glenzer2010,Zylstra2022}, laser-based colloid synthesis\cite{Zhang2017}, and material modifications\cite{Vorobyev2013}. 
Each of these applications would benefit greatly from thorough modeling capabilities that can reduce requirements for experimental tests, enhance analysis, or guide design. 
However, quantitative modeling requires knowledge of all energy transfer pathways and has to take into account the complex interplay of the different relaxation processes.

Recent results for copper\cite{Cho2011,Cho2016,Jourdain2018,Mahieu2018,Jourdain2021} show that models for the dynamic response after rapid energy deposition still lack predictive power. 
Indeed, these experimental data are highly controversial and show significant disagreements with standard modeling and theory\cite{Nguyen2023,Medvedev2020,Ng2024}.
One major source of uncertainty is the indirect determination of the temperature and time of the phase transition. 
We address this limitation by combining ultrafast electron diffraction experiments with atomistic simulations to investigate structural dynamics on the picosecond time scales where melting occurs. 
Ultrafast electron diffraction\cite{Mo2018} allows direct access to the dynamics of the phase transition on the atomic scale. 
Our data show clear signatures for structural changes and the analysis of the transient structure allows for the determination of the lattice temperature via the Debye-Waller factor. 
The latter gives access to the rate of electron-phonon energy transfer.
The experimental data are supplemented by Molecular Dynamics (MD) simulations for the copper nuclei coupled to heated and equilibrated electrons\cite{Maigler2024}. 
By using a highly optimized interatomic potential and an electron-phonon coupling strength validated by our experiments, we obtain good agreement with the measured melting kinetics including the onset, duration and completion of melting. 

\section*{Ultrafast electron diffraction}
\begin{figure}
\centering
\includegraphics[width=1\linewidth]{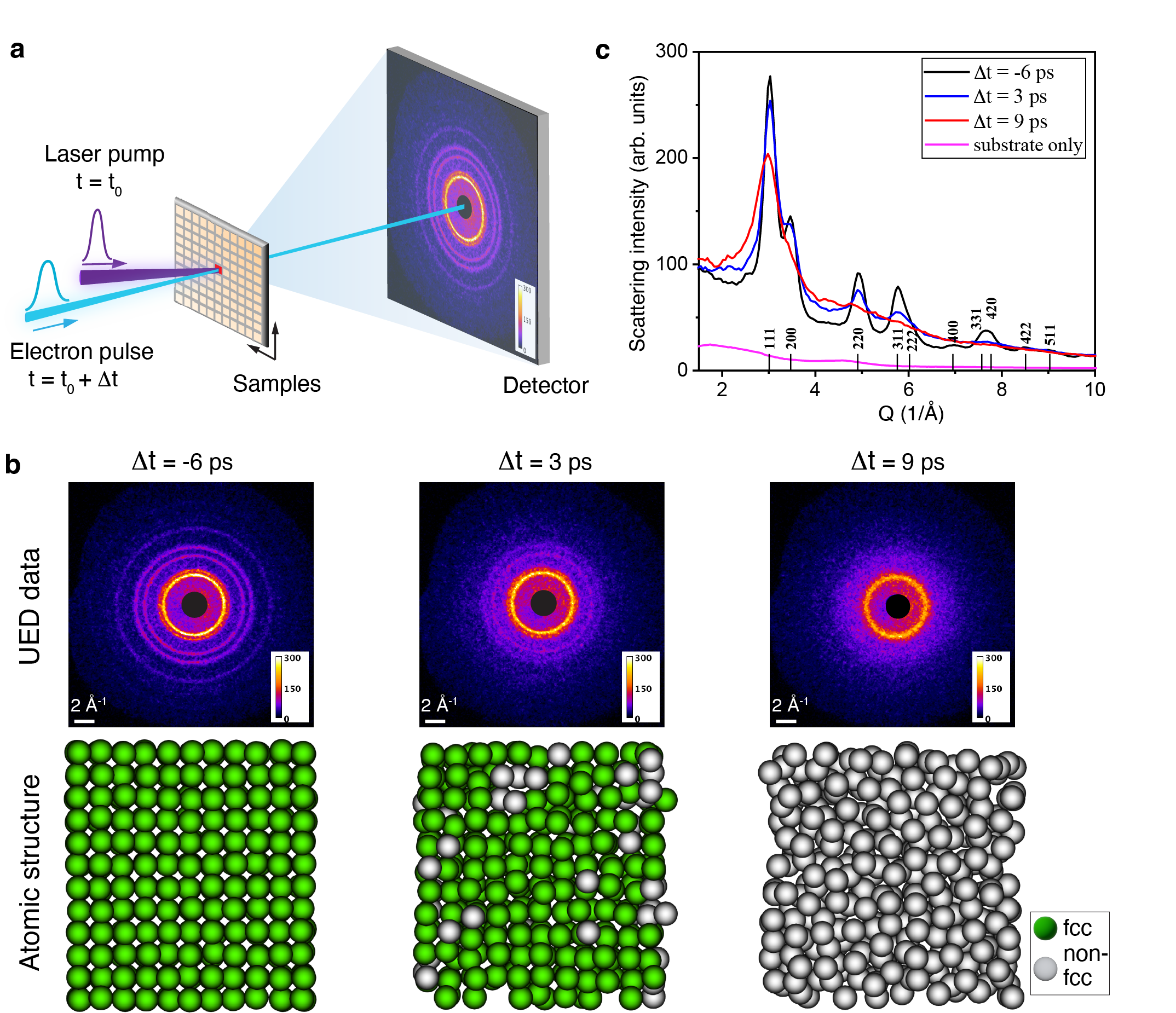}\caption{\linespread{1.0}\selectfont{} \textbf{Electron diffraction study of ultrafast melting of copper.} (a) Schematics of the experimental setup. The copper targets were pumped by intense \SI{130}{fs}, \SI{400}{nm} laser pulses and the induced solid-liquid phase transitions were probed with time-resolved electron diffraction using \SI{350}{fs} pulses of \SI{3.2}{MeV} electrons. (b) Top panel: sequence of diffraction patterns obtained at selected times over the solid-liquid phase transition obtained for an energy density of \SI{2.55}{\mega\joule\per\kilogram}. Bottom panel: snapshots of atomic configuration in a small subvolume as obtained by MD simulations to illustrate the transient atomic structures at corresponding time delays. Green spheres are atoms in the fcc phase, while grey spheres are non-fcc (liquid) atoms. (c) Radially averaged scattering intensity of the diffraction patterns shown in (b). The vertical black lines indicate the positions of diffraction peaks for fcc copper. For comparison, the intensity for only the unpumped Si$_3$N$_4$ substrate was included.
\label{fig:fig1} }
\end{figure}

Figure \ref{fig:fig1}a shows the schematic diagram of our experimental setup. 
We used \SI{40}{nm}-thick polycrystalline copper films deposited on \SI{30}{nm}-thick, free-standing, amorphous \ce{Si3N4} membranes as targets for our ultrafast melting study. 
We excited these targets from the copper side by \SI{130}{fs} long [full width at half maximum (FWHM)] laser pulses with a wavelength of \SI{400}{nm} and flat-top-like intensity profiles of $\sim$\SI{400}{\micro\meter} diameters. 
We expect uniform heating in the longitudinal direction because of the ballistic energy transport from the non-thermal electrons excited by the laser pulses, which have a mean free path of approximately 70 nm \cite{Cho2011}. 
The incident laser fluence during the measurement was set between \SI{105}{\milli\joule\per\square\cm} and \SI{213}{\milli\joule\per\square\cm},  corresponding to absorbed energy densities ranging from \(\epsilon = \SI[separate-uncertainty=true]{1.26\pm0.19}{\mega\joule\per\kilogram}\) to \(\epsilon = \SI[separate-uncertainty=true]{2.55\pm0.38} {\mega\joule\per\kilogram}\).
Thus, the energy densities in the target exceed the threshold of complete melting, that is $\sim$\SI[separate-uncertainty=true]{0.62}{\mega\joule\per\kilogram} (see Methods), by a factor of two to four.

To investigate the dynamic structural changes during the melting processes, we carried out ultrafast time-resolved electron diffraction measurements in a normal incidence transmission geometry, utilizing bunches of \SI{3.2}{\mega\electronvolt} (kinetic energy) electrons with bunch charges of $\sim$\SI{20}{fC}, and pulse durations of $\sim$\SI{350} {fs} (FWHM). 
We focused these relativistic electrons to diameters of $\sim$\SI{120 }{\micro\meter} on the targets. 
The scattered electrons were recorded by a phosphor-based detector located \SI{3.2}{m} downstream, providing a maximum momentum transfer $Q$ of $\sim$10$\,\angstrom^{-1}$. Here, $Q$ is defined as $Q = 4\pi \sin\theta/\lambda$ with $\theta$ and $\lambda$ representing the scattering angle and the de Broglie wavelength of the electrons,  respectively.  
At each pump fluence, we scanned pump-probe time delays (\(t\)) over several tens of picoseconds 
with \SI{0.5}{ps} time steps near time zero and larger intervals thereafter. 
At each time delay, we acquired and averaged a total of eight pump-probe events, resulting in diffraction patterns with excellent signal-to-noise ratios.

In Fig.~\ref{fig:fig1}b, we present a series of electron diffraction patterns taken before, during and after the solid-liquid phase transition for \(\epsilon= \SI{2.55}{\mega\joule\per\kilogram}\). 
At \(t = \SI{-6}{ps}\), far before the laser arrival, the diffraction pattern shows distinct Debye-Scherrer rings that are characteristic of the face-centered cubic (fcc) structure of polycrystalline copper. 
The radially averaged intensity lineout shown in Fig.~\ref{fig:fig1}c reveals a substantial background present below the diffraction peaks. 
This background predominantly arises from various scattering phenomena inside the copper film, including multiple elastic scattering and inelastic scattering events such as thermal diffuse scattering\cite{Siwick2004}.
In the case of a medium atomic-number material like copper, thermal diffuse scattering is expected to dominate the background given the relatively large elastic mean free path ($\sim\,$\SI{65}{\nano\meter}) of MeV electrons in copper. 

At \(t = \SI{3}{ps}\), the diffraction pattern displays a considerable qualitative change, marked by the vanishing of diffraction rings in the high-\(Q\) region and a reduction in the intensity of rings at low \(Q\). 
By contrast, the overall background signal is greatly enhanced. 
These observed changes are further illustrated by the scattering intensity lineout shown in Fig.~\ref{fig:fig1}c. 
It's worth noting that the diffraction pattern lacks a distinct feature of the liquid structure factor, which would otherwise broaden the width of the (111) peak and downshift its position in \(Q\), as shown by the intensity lineout taken at later times.
This suggests that the structure of the excited sample remains predominantly in a solid state at \(t = \SI{3}{ps}\). 
 
At a much later time, i.e., \(t = \SI{9}{ps}\), the pattern is characterized by a broad scattering ring near the (111) peak accompanied by a highly diffusive background. 
This is a signature of the liquid structure factor and implies that the sample has transitioned into a completely disordered state. Thus, melting is completed at this time.

\section*{Time-resolved electron scattering measurements}
To better understand the dynamics of melting and its dependence on energy density, we show the complete time history of the electron scattering signal obtained for two energy densities in Fig.~\ref{fig:fig2}.
We find that the evolution of the total scattering signal exhibits qualitative similarities for the two excitation conditions as they both feature: (i) an intensity decay of the Laue diffraction peaks, which is evident through the declining trend of the (220) peak; 
(ii) an increase of diffuse scattering signal spanning over the entire Q range, which is, however, more visible between diffraction peaks; 
(iii) the rise of the liquid scattering peak, with its primary peak position slightly below the (111) peak. 

To examine the first two effects, we select the signals of the (220) peak centered at \(Q = \SI{4.92}{\per\angstrom} \) and the diffuse scattering at \(Q = \SI{6.70}{\per\angstrom}\), and follow their time evolution. 
The results are shown in Figs.~\ref{fig:fig2}b and \ref{fig:fig2}e.
To quantify the dynamics of these transient signals, we fit the results with exponential functions defined by \(1 + A(1-\exp(-t/\tau))\) with \(A\) being the change of amplitude at steady state and \(\tau\) the time constant.
At \SI{2.55}{\mega\joule\per\kilogram}, we observe that the (220) peak decays on a time scale of \(\SI[separate-uncertainty=true]{2.9\pm0.2}{ps}\), as comparing to \SI[separate-uncertainty=true]{1.8\pm0.6}{ps} for the rise of the diffuse scattering signal.
For the lower energy density of \SI{1.26}{\mega\joule\per\kilogram}, we find slower changes with time constants of \SI[separate-uncertainty=true]{4.3\pm0.4}{ps} and
\SI[separate-uncertainty=true]{3.2\pm0.5}{ps} for the signals of the (220) peak and diffuse scattering, respectively.
It's crucial to recognize that under strongly driven conditions both lattice heating and structural phase changes contribute to the observed dynamics of the total scattering signal. 
While the diffuse scattering shows higher sensitivity, the polycrystalline nature of the sample prevents a clear interpretation of the underlying physical mechanisms.
In contrast, the decay of solid diffraction peaks allows to extract quantitative information on lattice heating and structural phase changes.  

The formation of the liquid peak during the melting process cannot be analyzed in the same way as it overlaps with solid diffraction peaks, particularly in the vicinity of the (111) peak where it is the strongest. 
This overlap makes it difficult to differentiate the two sources contributing to the local scattering intensity. 
To circumvent this issue, we instead track the width of the signal around the (111) peak, avoiding the need of decoupling the overlapping peak while still effectively characterizing the dynamics of liquid formation. 
This method relies on the principle that the overlapping peak width expands as crystallites decrease in size\cite{Xraybook} and the disorder effect intensifies\cite{Lin2010} during the melting process. 
The results are shown in Figs.~\ref{fig:fig2}c and \ref{fig:fig2}f for the two energy densities applied.   
We observe that the onset of width broadening exhibits a delay of $\sim$\SI{2}{ps} and $\sim$\SI{4}{ps} with respect to the excitation time.
We attribute this broadening to the onset of melting (\(\uptau_{\text{onset}}\)), corresponding to the time required to heat the lattice close to the melting temperature.  
A recent time-resolved X-ray diffraction study on ultrafast melting of polycrystalline palladium thin films found that laser-induced inhomogeneous strain would contribute to the (111) peak width broadening at energy densities near the melting threshold \cite{Antonowicz2024}. 
The onset of this effect, independent of excitation strength, was observed at 3 ps after laser excitation.
In our case, we cannot resolve this effect due to the limited momentum resolution of the electron scattering experiment. 
Besides, the clear dependence of \(\uptau_{\text{onset}}\) on excitation strength implies that laser-induced strain has a minor contribution in our case. 

We find that at the onset of melting \(\uptau_{\textrm{onset}}\), the lattice temperature is approximately \SI{1000}{K} for the stronger excitation (see Fig.~\ref{fig:fig4}b), 
amounting to $\sim$\SI{75}{\percent} of the nominal melting temperature ($T_{melt} = \SI{1358}{K}$ of copper\cite{CRCHandbook}). 
This suggests that the liquid nucleation is initiated by surface melting which occurs on the free surface and grain boundaries at temperatures below \(T_{melt}\)\cite{Jayanthi1985,Dash1999}.
For the lower energy density (\SI{1.26}{\mega\joule\per\kilogram}), it takes longer to reach the temperature sufficient for surface melting, hence resulting in a larger \(\uptau_{\text{onset}}\).
After melting has started, the width of the (111) peak is well described by exponential growth. 
In contrast to the decay of the (220) peak, both cases here exhibit similar time constants of approximately \(\SI[separate-uncertainty=true]{7.5\pm1.0}{ps}\), despite differences in the final equilibrated values at late delay times. 
As the width is connected to the rise of the liquid peak, the exponential form and the comparable time scales are indicative of homogeneous melting with similar liquid nucleation rates once the melting has started.

Our simultaneous measurement of the multiple orders of Laue diffraction peaks allows to quantify the lattice heating through the Debye-Waller factor (DWF)
\begin{equation}
\textrm{DWF}_{\textrm{hkl}}(t) = \frac{I_{\textrm{hkl}}(t)}{I_{\textrm{hkl}}^0} =  \exp \bigg[-\frac{1}{3}\Delta \langle u^2 \rangle(t)\cdot Q_{\textrm{hkl}}^2 \bigg],
\label{eq:DWF}
\end{equation}
where $I_{\textrm{hkl}}(t)/I_{\textrm{hkl}}^0$ denotes the peak intensity of reflection \((\textrm{hkl})\) at time \(t\) normalised to its value for a cold lattice and \(Q_{\textrm{hkl}}\) is the momentum transfer of this reflection.
This ratio is related to the time-dependent change of the mean square displacement of nuclei \(\Delta \langle u^2\rangle (t) \) which, in turn, is connected to the lattice temperature. 
Our data analysis of the DWF included the (111), (200), (220) reflection peaks, and the average of the (311) and (222) reflections, which are in close proximity. 
To minimize the effect of shot-to-shot variations in the electron bunch charge, we apply the DWF of the (111) peak as a reference to normalize higher-order DWFs (see Supplement Information). 
To additionally minimize the influence of hydrodynamic expansion and structural phase changes, we limit the DWF analysis to the initial \SI{4}{ps} after laser excitation which is half the time scale we estimate for the hydrodynamic response to become relevant (see Supplement Information).

We present the DWF results for an energy density of \(\epsilon\) = \SI{2.55}{\mega\joule\per\kilogram} in Fig.~\ref{fig:fig3}a, while the results for \(\epsilon\) = \SI{1.26}{\mega\joule\per\kilogram} are shown in the Extended Data Figure 1. 
When melting starts, the phase change lowers the fraction of solid phase in the heated film and accordingly results in an additional decay of the solid diffraction peak intensity. 
This effect is manifested by the emergence of the positive intercept with the linear fit to the logarithm of the DWF at time delays larger than \(\SI{2}{ps}\) as shown in Fig.~\ref{fig:fig3}b, which is not expected if it were a sole DWF effect\cite{Mo2018RSI}. 
The time when the deviation starts is also consistent with the onset of melting determined from the width of the (111) peak. 
At the lower energy density (\(\epsilon\) = \SI{1.26}{\mega\joule\per\kilogram}), the positive intercept appears at a later time due to the delay in the onset of the melting process, c.f. Extended Data Figure 1b.  

We determine the ion temperature evolution using the measured DWFs and the mean square displacement of copper that accounts for the anharmonic contribution\cite{Day1994,Shepard2000} (see Supplementary Information). 
For the strong excitation shown in Fig.~\ref{fig:fig3}c, the ion temperature increases rapidly from the room temperature to $\sim$\SI{1750}{K} at \(t = \SI{4}{ps}\).
This corresponds to an average heating rate of \(\sim \SI{3.6e14}{\kelvin\per\second}\), implying that the system enters into a regime where homogeneous melting dominates the phase transition\cite{Lin2006,Mo2018}, consistent with our interpretation for the width of the rising liquid peak.

Importantly, the measured ion temperature evolution allows to test models for the electron-lattice coupling (\(G_{ei}\)) in copper. 
We apply the two-temperature model (TTM) to describe the ion temperature evolution in fs laser-heated copper (see Methods), and compare the results with the experimental data. 
We test two different expressions for \(G_{ei}\) and show the comparison in Fig.~\ref{fig:fig3}c. 
The coupling suggested by Lin \emph{et al.} \cite{Lin2008} results in lattice temperatures clearly exceeding the measured data, while the model by Migdal \emph{et al.} \cite{migdal2016} predicts a more gentle increase which aligns well with the experimental results.
The same observation is also found for the weaker excitation 
(c.f.\ Extended Data Figure 1c). 
When calculating DWFs with ion temperature from TTM simulations using \(G_{ei}\) from Migdal \emph{et al.}, we find good agreement with the experimental data for all diffraction peaks and both excitation strengths (see Fig.~\ref{fig:fig3}a and Extended Data Figure 1a). 

\section*{TTM-MD simulations}
We conduct TTM-MD simulations of ultrafast melting in copper using the LAMMPS code \cite{thompson2022} (see Methods) to understand the transient atomic structures  during the melting transition. 
We select a highly optimized embedded atom method (EAM) potential developed by Sheng \emph{et al.}\cite{sheng2011} to describe the mutual interaction between copper atoms.
The initial photo-excitation is modeled as a sudden increase in the electronic temperature. 
For the energy transfer between the lattice and electrons, we select the model by Migdal \emph{et al.} that matches the measured lattice temperatures (see Fig.~\ref{fig:fig3}c). 

Figure \ref{fig:fig4} presents the simulation results for the time evolution of the average ion temperature, \(T_i\), and the fraction of fcc atoms, \(\psi\), for \(\epsilon\) = \SI{2.55}{\mega\joule\per\kilogram}. Additionally, snapshots of atomic configurations at different times sampling the solid-to-liquid phase transition are shown. 
Due to our choice of the electron-lattice coupling, the MD simulation follows the measured lattice temperature for the first \SI{4}{ps}.
Similarly, the simulation results for the case of \(\epsilon\) = \SI{1.26}{\mega\joule\per\kilogram} are shown in the Extended Data Figure 2.

The atomic configurations indicate that the film remains nearly crystalline up to \SI{2}{ps} after excitation, at which time \(T_i \approx 1025\) K. 
Shortly afterwards, but before the nominal melting temperature is reached, we observe the onset of melting, marked by a weak decrease in \(\psi\). 
The snapshots show that this decrease in order stems mainly from the free surface at the top of the sample. 
The transition to atoms not in a fcc lattice strongly accelerates once the melting temperature is reached at roughly \SI{3}{ps}. 
From this time onward we observe that the entire volume contributes to the phase transition in agreement with the prediction of the homogeneous melting regime. 
At \SI{4}{ps}, an ion temperature of \(T_i \approx 1667\) K is reached, and roughly half the sample has transitioned to the liquid phase. 
Although we have slightly more disordered atoms near the free surface, this part does not contribute substantially to the overall melting process.  
In particular, we do not observe a melting wave from the surface but a transition to the disordered phase homogeneously distributed throughout the sample once the melting temperature is reached. This process is almost finished around \SI{6}{ps}.

Due to the fast heating rate, the ion temperature rises steadily far beyond the melting temperatures and also exceeds a proposed stability limit of the lattice\cite{Arefev2022}, that is $\sim$1.25\,$T_{melt}$, \SI{4.1}{ps} after the excitation. 
At this limit of superheating, a rapid collapse of the crystalline structure is predicted.
Interestingly, we do not observe this behaviour. 
In our simulations, homogeneous melting starts earlier and progresses smoothly without any acceleration once this proposed superheating limit is exceeded. 
Indeed, we only observe changes in slope of the transition rates at \SI{2}{ps} when melting starts in the entire volume and after \SI{6}{ps} when almost all atoms have been converted.
To further investigate this behavior, we examine local atomic configurations near the center of the film after the superheating limit (see Fig.~\ref{fig:fig4}c). 
We find that small pockets of liquid appear spontaneously from fluctuations at random positions which then grow further and drive the melting process. 
These findings point to a smooth transition starting in the bulk at the nominal melting temperature and progressing on a time scale which is defined by the internal nucleation and growth dynamics. 

\section*{Discussion and Perspective}
\begin{figure}
\centering
\includegraphics[width=1\linewidth]{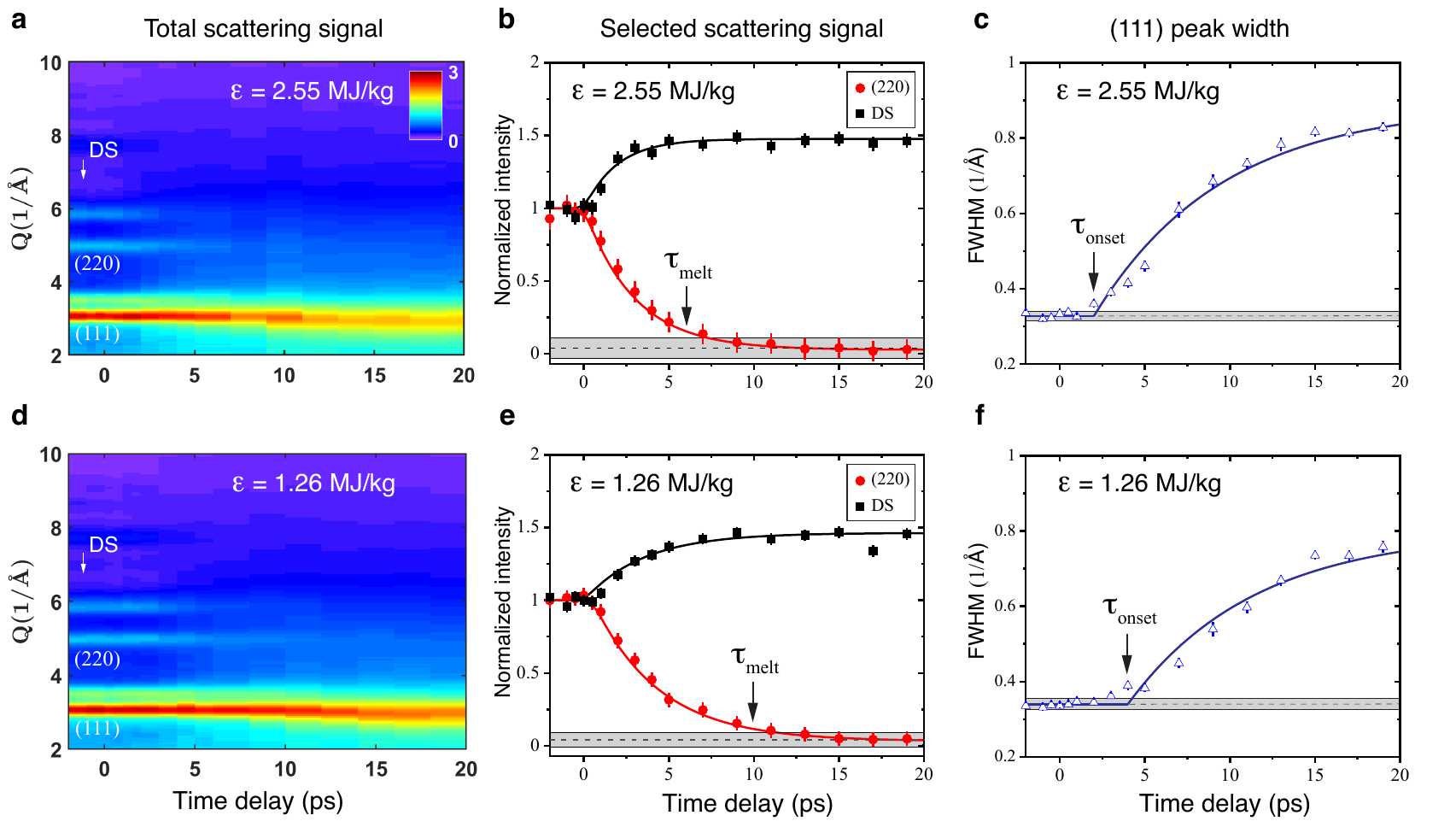}
\caption{\linespread{1.0}\selectfont{}\textbf{Temporal evolution of the electron scattering signal.} The upper panels are results for an energy density of \SI{2.55}{\mega\joule\per\kilogram}, the lower panels are for \SI{1.26}{\mega\joule\per\kilogram}. (a) and (d) Time-resolved result of the total scattering signal. (b) and (e) Evolution of the intensity of the decaying (220) peak (red spheres) and the rise of diffuse scattering signal at Q = 6.8\,\AA$^{-1}$ (black squares). Error bars represent one standard deviation (SD). The gray band represents the ground intensity level (mean value $\pm$ 1 SD) reached at late times.
The vertical arrow marks the time of complete melting \(\mathrm{\uptau_{melt}}\). To account for the relatively large time intervals at late times, we define \(\mathrm{\uptau_{melt}}\) as half of the time interval (1 ps) preceding the drop in (220) intensity to the liquid scattering level. (c) and (f) Time-resolved width of the (111) peak (FWHM) to monitor the signal from liquid scattering developing during the melting process. The gray band represents the initial constant level of the (111) peak width (mean value $\pm$ 1 SD). The onset of changes in the width  (\(\mathrm{\uptau_{onset}}\)) is marked by the vertical arrow.
\label{fig:fig2} }
\end{figure}

\begin{figure}
\centering
\includegraphics[width=1\linewidth]{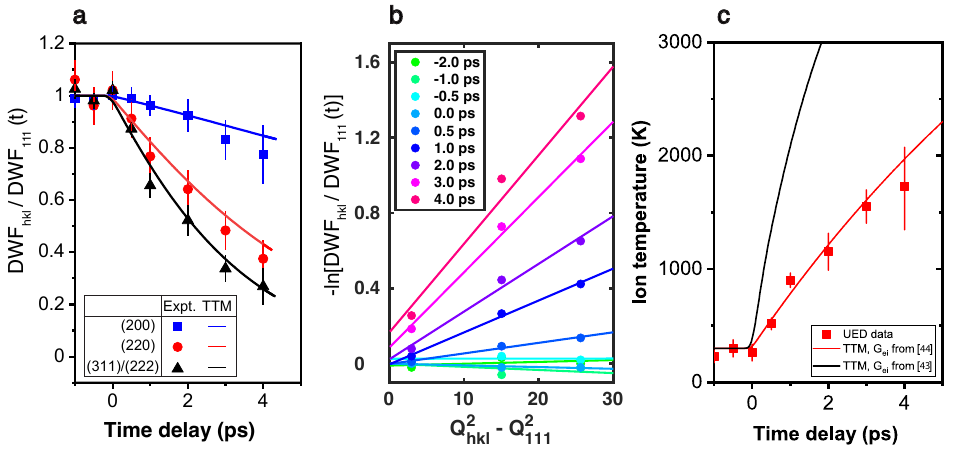}
\caption{\linespread{1.0}\selectfont{}\textbf{Determining the lattice heating}. (a) Experimental data of the decay of normalized diffraction peaks, normalized to the (111) peak (discrete data points), in comparison with corresponding DWFs that are calculated from TTM simulation results (solid lines). In these TTM simulations, we adopted a temperature-dependent \(G_{ei}\) from Migdal \emph{et al.}\cite{migdal2016} (see Methods). (b) Negative logarithm of the normalized DWF (discrete data points) as a function of \(Q_{hkl}^2-Q_{111}^2\), overlaid with linear fits (solid lines) to extract the mean displacement \(\Delta \langle u^2\rangle (t)\). (c) Evolution of lattice temperature derived from the measured DWFs compared to TTM simulation results based on electron-ion coupling according to Lin \emph{et al.}\cite{Lin2008} (black line), and Migdal \emph{et al.}\cite{migdal2016} (red line). 
Data in all graphs are for \(\epsilon\) = \SI{2.55}{\mega\joule\per\kilogram}.
\label{fig:fig3} }
\end{figure}

To provide further insight into the melting dynamics, we examine the energy density dependence of characteristic times associated with ultrafast melting, i.e., the times for melt onset and melt completion as obtained from both electron diffraction measurements and TTM-MD simulations.  
Experimentally, we define melt onset as the moment when the (111) peak width starts to broaden, and melt completion as the point when the (220) peak disappears (c.f.\ data in Fig.~\ref{fig:fig2}).
In the simulations, the gradual nature of the phase transition requires a statistical definition of these times. 
Adapting the approach by Arefev \emph{et al.}\cite{Arefev2022}, we define melt onset as the time when 98-95\% of the atoms are in the fcc lattice, and melt completion as the time when this fraction falls in the range of 5-2\%. 

Figure \ref{fig:fig5} demonstrates that the measurements are well reproduced by our TTM-MD simulations for both characteristic times.  
We also observe that all time scales exhibit a gentle decrease when the target absorbs more energy.   
This trend is consistent with the results for ultrafast melting of gold within the homogeneous melting regime\cite{Mo2018}.
However, copper melts slightly faster than gold (10-20\,ps for the excitation
strength shown) owing to its larger electron-phonon coupling strength.

\begin{figure}
\centering
\includegraphics[width=1\linewidth]{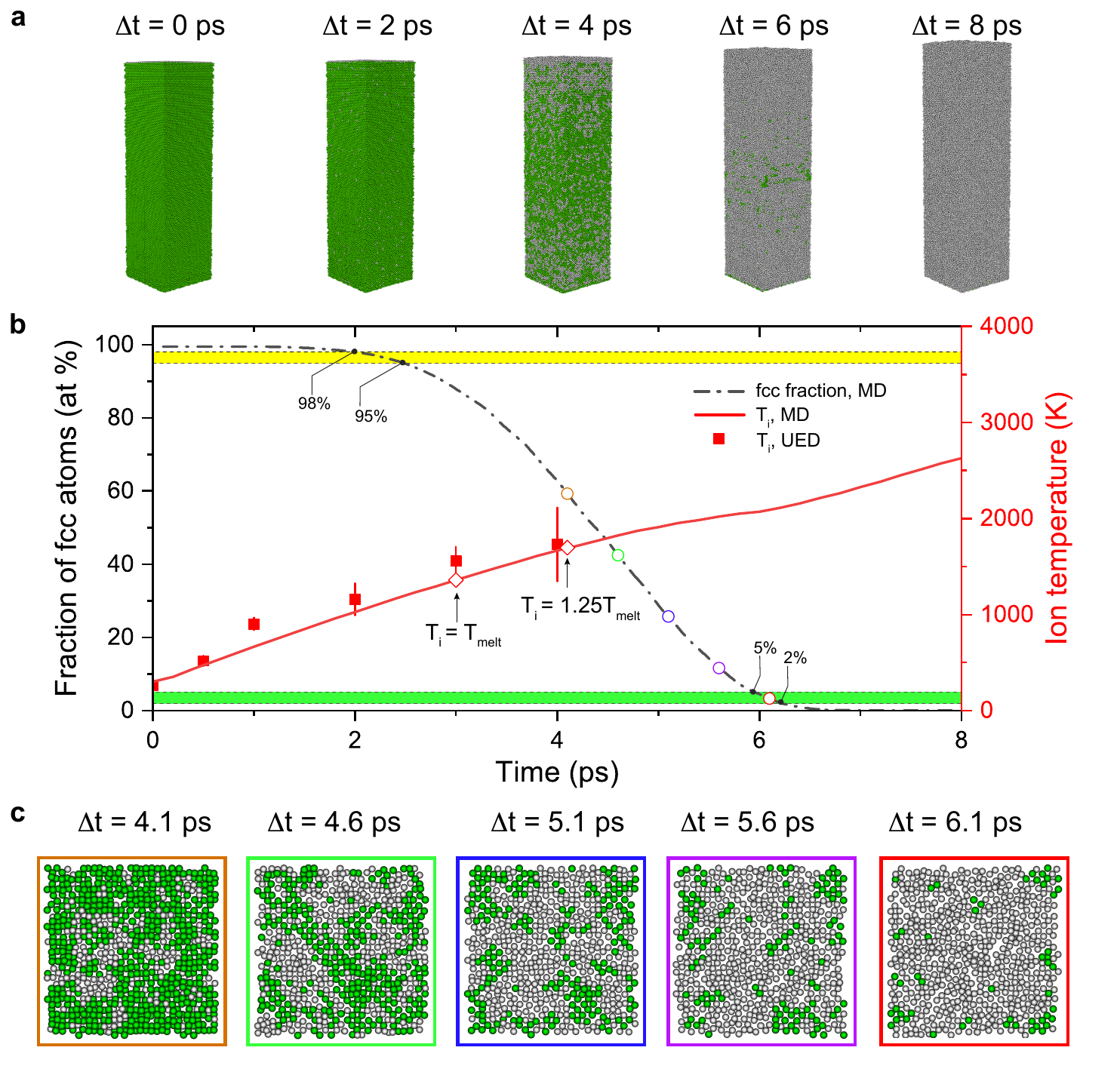}
\caption{\linespread{1.0}\selectfont{}\textbf{Results of TTM-MD simulation.} (a) Snapshots of atomic configurations for different times distributed over the melting transition. The top side of the slab is free to move while the bottom side is fixed to simulate the substrate. The atoms in the simulations are colored according to the surrounding atomic structure: green are atoms in a fcc lattice, grey are disordered (liquid) atoms. 
(b) Evolution of the fraction of fcc atoms (dash-dot line) and the average ion temperature (red solid line). The yellow band highlights fcc fractions ranging from 98\% to 95\% (onset of melting); the green band fractions from 5\% to 2\% (melt completion). The red squares with error bars label lattice temperatures from the electron diffraction measurements for comparison. The energy density considered here is \SI{2.55}{\mega\joule\per\kilogram}.
(c) Snapshots of atomic configurations after reaching the predicted superheating limit of 1.25 \(T_{melt}\) for selected areas at the center of the slab. The time points for the snapshots are marked out with color-coded circles on the fcc fraction curve in (b). 
\label{fig:fig4} }
\end{figure}

\begin{figure}
\centering
\includegraphics[width=0.7\linewidth]{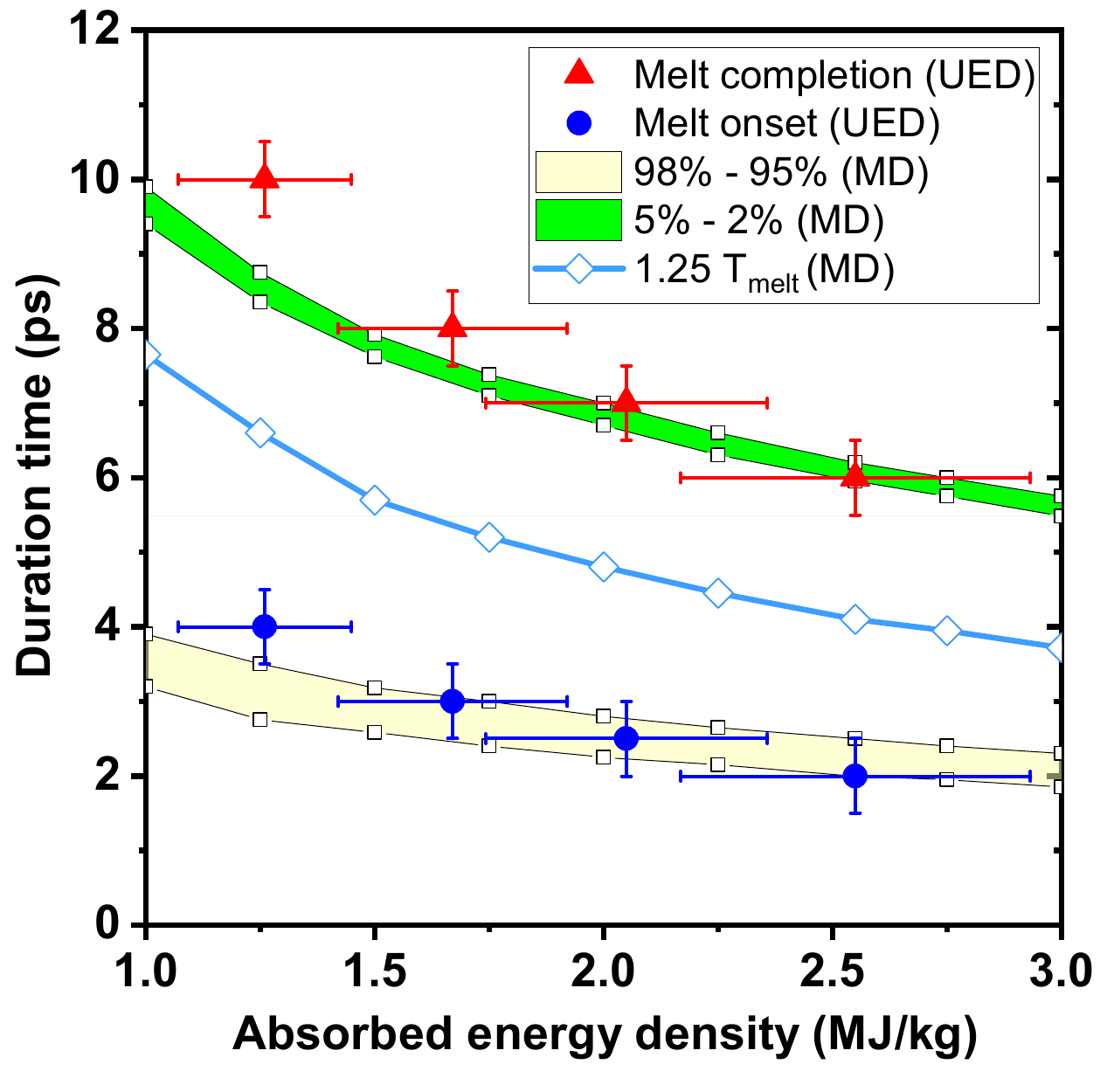}
\caption{\linespread{1.0}\selectfont{}\textbf{Energy density dependence of melting times for copper.} The measured times for melt completion and onset are represented by red triangles and blue circles, respectively. Error bars represent one standard deviation of statistical uncertainties. The data are compared to results from TTM-MD simulations, with open squares indicating the energy densities that have been modeled. The green shaded area corresponds to times when the fraction of fcc atoms is between 5\% and 2\% (melt completion), the yellow shaded area for times with fcc fractions from 98\% to 95\% (onset of melting). The light blue line with open diamonds represent the times when the average ion temperature reaches \(1.25\,T_{melt}\).
\label{fig:fig5} }
\end{figure}

We also observe that, in all cases, the onset of melt is reached significantly earlier than the superheating limit of 1.25\,$T_{melt}$ for both simulations and experiments.
Indeed, the simulation results show that \(T_i\) is just slightly below the nominal melt temperature \(T_{melt}\) when the material starts to melt (c.f.\ Fig.\ref{fig:fig4}), consistent with the surface melting effect observed in electron diffraction measurements. 
Moreover, we don't observe any special features, like a kink in the transition rates, around 1.25\,$T_{melt}$ for all cases investigated although this predicted superheating limit is always crossed  between the onset and completion of melting.

We can now construct a physical picture of the energy flow 
in copper driven by intense short-pulse laser excitations.  
As an initial condition, the laser pulse deposits the energy to the electrons, which are thermalized within the pulse duration\cite{Mueller2013PRB}.
Under our experimental conditions, the initial temperatures of the electrons are approximately two orders of magnitude higher than the ion temperature, forming highly non-equilibrium conditions.
The subsequent energy transfer from the electrons to the lattice is determined by the strength of electron-ion coupling, for which strongly deviating predictions have been made\cite{Lin2008,migdal2016,Medvedev2020,Ng2024}. 
In this study, we track the lattice temperature at early times using the Debye-Waller effect, which dominates the intensity decay of solid diffraction peaks.
We find that weak electron-phonon coupling as suggested by Migdal \emph{et al.}\cite{migdal2016} agrees well with the experimental temperature evolution whereas the models predicting a stronger coupling disagree with the data significantly. 
This finding is further supported by the ion temperatures in the liquid formed after melting, c.f. Supplementary Fig. S1, determined by comparing the measured liquid structure factor with density functional theory (DFT)-MD simulations. 

To interpret the melting dynamics, one has to consider the fact 
that the relatively strong electron-ion coupling in copper leads to rapid lattice heating, allowing the crystal to reach the melting temperature and the superheating limit within a few picoseconds (c.f.\ Fig.~\ref{fig:fig5}).
It is the prevailing view that melting within a superheated crystal occurs as a threshold-like process, marked by the rapid and widespread homogeneous nucleation of liquid regions\cite{Arefev2022}.  
Contrary to this prediction, our measurements and TTM-MD simulations clearly reveal that melting in the thin copper films begins much earlier slightly below the nominal melting temperature. 
This fact cannot be fully explained by the reduced superheating limit resulting from the dynamic relaxation of stresses generated by the ultrafast laser heating\cite{Ivanov03PRL,Arefev2022}.
Instead, the observed melting onset is likely triggered by melting starting at free surfaces and grain boundaries.
However, heterogeneous melting only plays a minor role in the very beginning of the phase transition and the dominant contribution comes from homogeneous melting that starts at the melting temperature and smoothly transitions over the superheating limit. 
Moreover, our TTM-MD simulations, despite not considering phonon hardening\cite{Ernstorfer2009,Descamps2024}, still reproduce both the early-time lattice temperature evolution and the complete melt time. 
This observation suggests that phonon hardening only plays a minor role under the present experimental conditions.

Previous work by Jourdain \emph{et al.}\cite{Jourdain2021} reported characteristic times less than \SI{1}{ps} for the loss of lattice periodicity in fs laser-heated copper. These results were obtained at comparable energy densities using the technique of X-ray absorption near-edge spectroscopy (XANES).  
Under the driven conditions, homogeneous melting is the primary mechanism and hence the difference in sample thickness between the XANES measurement (80\,nm) and the current UED experiment (40\,nm) cannot account for the discrepancy. 
Instead, this apparent contradiction reflects the different observables of the two techniques: the transition of electronic structure probed by XANES is more pronounced during melting than that of ionic structure probed by diffraction\cite{Jourdain2021}.
These results underscore the importance of directly observing atomic structure and temperature when investigating the dynamics of melting. 

Our electron diffraction measurements provided an atomic-level view of ultrafast melting in copper driven by intense femtosecond laser excitations.  
The pump-probe scheme delivered highly detailed structural information with subpicosecond time resolution, capturing lattice heating and loss of long-range structure as the material transitioned from the cold solid to the liquid state.
Through the DWF, we determined the ion temperatures at early times when the target is mostly solid and, thus could test models for the electron-phonon coupling.   
We found good agreement with a relatively slow energy exchange rate in copper. 
Real-time observation of the structural evolution revealed an onset of melting slightly below the nominal melting point, followed by a much faster transition once the melting temperature is reached. 
We attributed the first effect to melting at free surfaces and grain boundaries
and the acceleration to the initiation of melting throughout the volume of the sample. 
Our TTM-MD simulations of the ultrafast melting dynamics showed very good agreement with the diffraction data, in particular with respect to the energy density-dependence of melt onset and completion. 

Our study provides crucial insights for refining theories of ultrafast melting in metals.
In particular, the lack of evidence for a collapse of the lattice structure at the proposed superheating limit challenges the prevailing view that such a collapse should be triggered at this temperature.
These results for copper can have strong implications for other metals and advancing materials processing techniques that aim to achieve atomic-level precision. \\

\begin{methods}
\subsection{Experimental details.}
Our ultrafast pump-probe experiment was performed at the MeV-UED instrument of the Linac Coherent Light Source (LCLS) user facility at SLAC National Accelerator Laboratory~\cite{Weathersby2015,Mo2016}.
The MeV electrons for diffraction were generated by a Linac Coherent Light Source (LCLS)-type photocathode radio-frequency (RF) gun.
The RF gun is powered by a pulse-forming-network based modulator and a 50-MW S-band klystron.
Time-resolved diffraction experiments were performed at normal incidence in transmission geometry with relativistic electrons with a kinetic energy of 3.2 MeV. 
The relativistic electrons were focused by two separated solenoids installed after the RF gun onto the target with diameters of $\sim\! \SI{120}{\micro\metre}$ (FWHM), bunch charges of $\sim\! \SI{20}{fC}$ and pulse durations of $\sim\! \SI{350}{fs}$ at FWHM. 
The electron detector was located \SI{3.2}{m} away from the sample and consisted of a P43 phosphor screen, a lens system and a sensitive Electron-Multiplying CCD (EMCCD) camera. 
In the middle of the phosphor screen, a \SI{1.6}{mm}-diameter through-hole  prevented the zero-order diffraction signal from saturating the CCD image at high gain during the experiment. 

The samples employed in this study were \SI{40}{nm}-thick polycrystalline copper films coated on 30-nm-thick, free-standing, amorphous \ce{Si3N4} membranes. 
These samples were mounted on an array of \SI{350}{\micro\metre} \(\times\) \SI{350}{\micro\metre} windows supported by a 2" \(\times\) 1" Si wafer. 
The samples were uniformly excited from the copper side by focused \SI{400}{nm}, \SI{130}{fs} (FWHM) laser pulses at $\sim 4^{\circ}$ angle of incidence with a flat-top-like intensity profiles of $\sim \SI{400}{\micro\metre}$ diameters.
The flat-top pump profile was achieved with 10:1 imaging of an aperture outside the target chamber by a single lens with a focal length of 25 cm. 
The root-mean-square intensity variation was approximately 15$\%$ of the averaged value within the full probed area. 
During our melting study, the pump-probe measurements were conducted in a range of incident laser fluences between \SIrange{105}{210}{\milli\joule\per\centi\metre\squared}. 
In this fluence range, the absorption coefficient of 400 nm laser light was measured to be constant at $\sim$43\% in an offline reflection and transmission experiment.

The different excitation states of copper samples were reached by changing the absorbed energy densities $\epsilon$, which is defined as the amount of absorbed laser energy over the mass of the excited material and is given by: $\epsilon=(E_{inc}\times A)/(S \times d \times \rho)$, where $E_{inc}$ is the measured incident laser pulse energy within the probed area $S$, $A$ is the measured absorption coefficient of the incident laser energy for our samples, $d$ is the copper layer thickness, and $\rho$ is the mass density for copper. 
We note here that the thermal energy loss to the substrate is neglected as a first-order approximation due to the short delay time window relevant to this melting study. 

\subsection{Complete melting threshold.}
The complete melting threshold is defined as the amount of energy that is required to heat the material from room temperature to its equilibrium melting temperature and to melt the entire film at the melting temperature.  
This complete melting threshold can be expressed in terms of absorbed energy density, $\epsilon_m$, given by :
\begin{equation}
\epsilon_{m} =  \int_{300}^{T_m}C_i\, dT + \Delta H_m
\label{CMT}
\end{equation}
where $C_i$ is the ion specific heat capacity, and $\Delta H_m$ is the enthalpy of fusion. 
Note that the electronic contribution to $\epsilon_m$ is very small (\(\sim 1\%\)) and hence it is neglected here.
Using the following parameters: $\Delta H_m$ = \SI{208.7}{\kilo\joule\per\kilogram},  $T_m$ =1358\,K \cite{CRCHandbook}, and \(C_i =\) \SI{3.5e6}{\joule\per\meter^3\per\kelvin}, which was obtained by the Dulong-Petit law, we estimated $\epsilon_m$ to be approximately \SI{0.62}{\mega\joule\per\kilogram} for copper.

\subsection{Debye-Waller Factor and two-temperature model.}
The attenuation of the diffraction peak intensity can be modeled using the DWF and the temperature evolution results of the TTM simulations. 
In the TTM framework, electrons and ions are treated as two distinct subsystems, each defined by its own temperature. 
Thermal equilibrium between the two subsystems is governed by the electron-ion coupling strength $G_{ei}$. 
Given that uniform heating is expected under our experimental conditions, electron thermal conduction can be neglected and the model equations are given by: 
\begin{subequations}
\begin{ceqn}
\begin{eqnarray}
C_e(T_e)\frac{\partial T_e}{\partial t}&=&-G_{ei}(T_e-T_i)+S(t)\\
C_i\frac{\partial T_i}{\partial t}&=&G_{ei}(T_e-T_i)
\label{eq:TTM}
\end{eqnarray}
\end{ceqn}
\end{subequations}
where $T_e$ and $T_i$ are the electron and ion temperature, $S(t)$ is the temporal profile of the absorbed laser pulse, which is a Gaussian function with FWHM of \SI{130}{fs}.
The electron heat capacity, \(C_e(T_e)\), was calculated from the electronic DOS and the distribution functions\cite{Maigler2024}.
The ion heat capacity, $C_i$, was assumed constant at \SI{3.5e6}{\joule\per\meter^3\per\kelvin}, which was obtained by the Dulong-Petit law.

With the simulated temperature evolution, the DWF at given \(t\) and \(Q\) can be calculated by expression \eqref{eq:DWF} which has been described in detail elsewhere \cite{Mo2018RSI}. 
Several studies have quantified the temperature dependence of the mean square displacement for copper. 
In the Supplementary Information, we provide a detailed comparison of these data and assess their impact on the derived $T_i$ values. 

\subsection{TTM-MD simulations.\label{md-simulation}} 
We carried out atomistic simulations using the approach of the Two-Temperature Model coupled to molecular dynamics simulations for the nuclei (TTM-MD) that is available in the LAMMPS (Large-scale Atomic/Molecular Massively Parallel Simulator) code\cite{thompson2022}. 
The TTM-MD approach adopted in our simulations is based on the work by Duffy and Rutherford\cite{Duffy2007,Rutherford2007} and has been developed further to account for adaptive electron background voxel interactions and temperature dependent coupling parameter\cite{Maigler2024}. 

In the TTM-MD framework, TTM simulates the energy exchange between electronic and atomic subsystems, while MD models the behavior of the heavy particles. The electronic system, modeled as a continuum overlapped with the MD ensemble, was represented using the following heat diffusion equation:
\begin{equation}\label{eq:heat_diffusion}
    \centering
    C_e\left(T_e\right)\frac{\partial{T_e}}{\partial{t}} = \nabla\cdot\left(\kappa{_e}\nabla{T_e}\right)-G_{ei}\left(T_e-T_i\right) + S\left(z,t\right),
\end{equation}
where $\kappa_e$ is the electronic heat conductivity,  and $S\left(z,t\right)$ is a source term that represents the energy deposited by the laser. 
To align with the experimental conditions, our simulations assume a uniformly heated target with an electron temperature defined by the deposited energy density as the initial state.
Consequently, the source term $S\left(z,t\right)$ can be omitted.

The equations of motion solved by the MD part is 
\begin{equation}\label{eq:newton}
    \centering
    m_i\frac{\partial{\textbf{v}_i}}{\partial{t}} = \textbf{F}_i -\gamma\textbf{v}_i+\textbf{f}_L\!\left(T_e\right) \,.
\end{equation}
Here, $\textbf{v}_i$ is the velocity of atom $i$ with mass, $m_i$. $\textbf{F}_i$ is the classical force acting on the atom calculated using a highly optimized EAM potential developed by Sheng \emph{et al.}\cite{sheng2011}, $\gamma$ represents a frictional drag force, $\textbf{f}_L$ is the stochastic force. 
The friction and stochastic terms allow energy to be added/removed from the ions to represent energy transfer with electrons, hence $\gamma$ is related to the electron-phonon coupling strength via
\begin{equation}\label{eq:gamma}
    \centering
    \gamma\ = \left(\frac{V}{N^\prime}\right)\frac{m}{3k_B} \, G_{ei}, 
\end{equation}
where $V$ is the volume of a coarse grain ionic voxel and $N^{\prime}$ is the number of atoms in the voxel.

The simulated copper slab has a dimension of \qtyproduct{40 x 10 x 10}{nm}, comprising a total of \SI{1.35E5}{copper} atoms. 
The simulation box measures \qtyproduct{150 x 10 x 10}{nm} and employs periodic boundary conditions in each spatial direction. 
In order to simulate effects of substrate interaction and to avoid unphysical  pressure reflections of stiff harmonic potentials fixed at the lower side of the slab \cite{Maigler2024}, 10 layers of substrate atoms are kept at a constant temperature of \SI{300}{K} (NVT ensemble) but allowed to interact with the copper slab through their force potential. 
Below this region another region of 5 ghost atom layers are set which are rigidly kept at their initial position and may only exert an influence on the substrate region through force potential interaction \cite{ZHOU2022}. 
Using this approach, unphysical pressure or (shock) wave reflections are dampened out.
The opposite side of the slab is left free to expand into vacuum.  

The system was first equilibrated for \SI{1}{ns} at \SI{300}{K} employing  \emph{NVT} ensemble.
The equilibrated configuration was then employed as the initial condition for the TTM-MD simulations conducted under \emph{NVE} ensemble. 

Identification of fcc and non-fcc atoms was performed by means of the  compute/ptm package, which is based on the polyhedral template matching algorithm (PTM) \cite{Larsen2016}. 
We employed a RMSD cutoff parameter of 0.15 and considered output of fcc, hcp, bcc, ico, and sc structures.
In the present study, since the fraction of crystalline structures other than fcc was found to be insignificant (maximum of few tens of atoms compared to the rest of the ensemble), we can safely attribute the non-fcc copper atoms as liquid atoms in our simulations.

\subsection{Data availability.}
The datasets generated and analyzed during this study are available from the corresponding author upon reasonable request.

\end{methods}

\bibliography{UED_Cu}

\begin{addendum}

 \item Use of the MeV-UED instrument, operated as part of the LCLS, at SLAC National Accelerator Laboratory, is supported by the U.S. Department of Energy (DOE), Office of Science, Office of Basic Energy Sciences under Contract No. DE-AC02-76SF00515.
 M.Z.M. acknowledged the support by DOE Fusion Energy Sciences under FWP $\#$101242, and U.S. DOE, Laboratory Directed Research and Development program at SLAC National Accelerator Laboratory, under contract DE-AC02-76SF00515.
 S.H.G. acknowledged the support by DOE Fusion Energy Sciences under FWP $\#$100182, and under FWP $\#$100866.
 A.B. acknowledges the support by the north german alliance for supercomputing NHR.
 K.S.T. acknowledges financial support by the Deutsche Forschungsgemeinschaft (DFG, German Research Foundation) through Project No. 278162697-SFB 1242.
X.J.W. acknowledges support by the Deutsche Forschungsgemeinschaft (DFG, German Research Foundation) through the Collaborative Research Centre (CRC) 1242 (project No. 278162697, project C01 Structural Dynamics in Impulsively Excited Nanostructures) \& the funding by the DFG Germany's Excellence Strategy - EXC 2033 - 390677874 – RESOLV.
\item[Author Contributions] M.Z.M. conceived and designed the study. M.Z.M., B.K.O., K.S.T., Z.C., R.K.L., X.S., X.J.W., and S.H.G. performed the experiment, M.Z.M. analyzed the UED data and performed the TTM simulations, M.B.M. and J.S. performed and analyzed the TTM-MD simulations, A.B. and R.R. performed and analyzed the DFT-MD simulations, M.Z.M., T.H., D.O.G. K.S.T. and B.R. contributed to the interpretation of the results, M.Z.M., M.B.M., D.O.G. and B.R. wrote the manuscript with revisions from all the authors.
 
 \item[Competing Interests] The authors declared that they have no
competing interests.

 \item[Correspondence] Correspondence and requests for materials
should be addressed to \\
M.Z.M.~(mmo09@slac.stanford.edu)
\end{addendum}

\newpage

\newpage

\newpage
\newpage

\newpage

\clearpage
\begin{extendedfigure}
\centering
\includegraphics[width=0.95\linewidth]{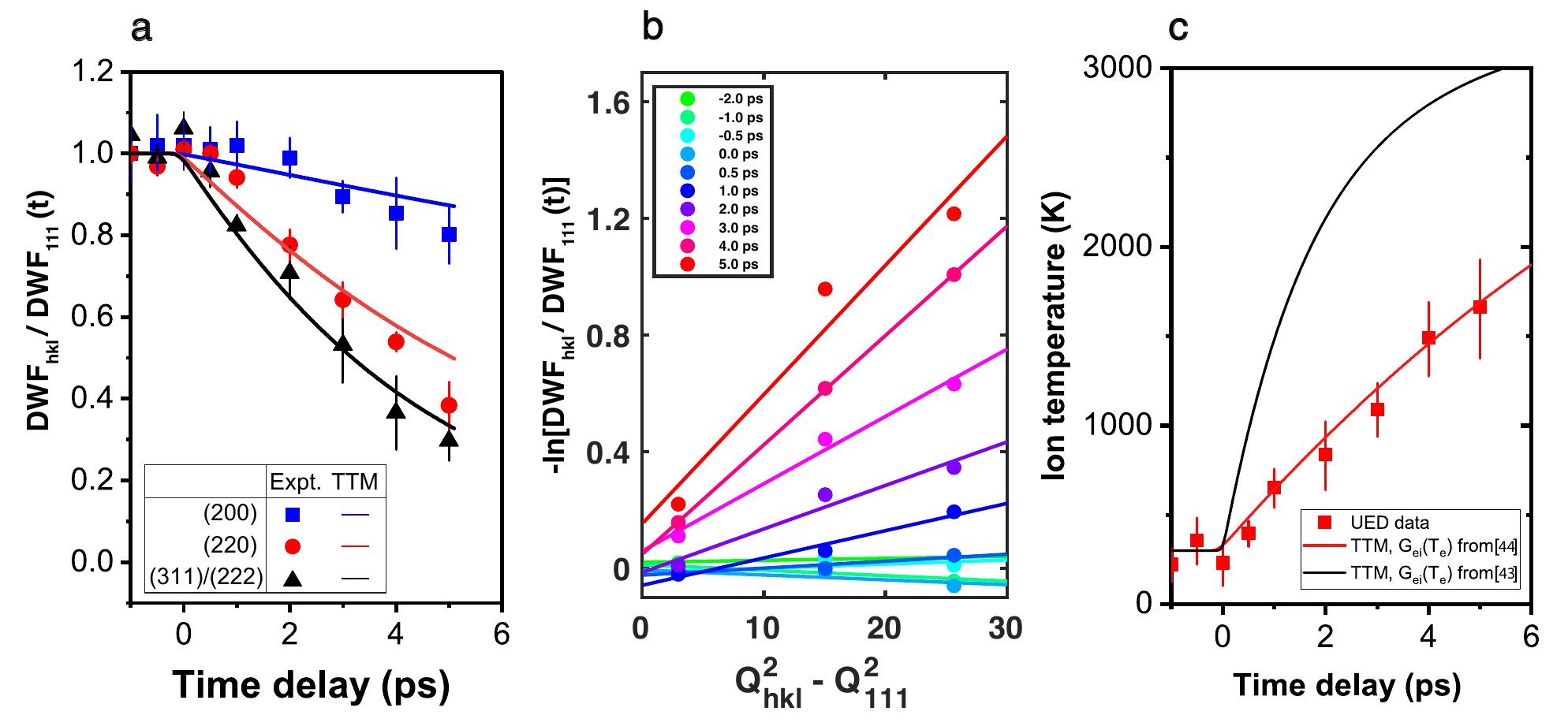}
\caption{\linespread{1.0}\selectfont{}\textbf{UED measurement of lattice heating in fs laser-heated copper.} The results are for \(\epsilon\) = \SI{1.26}{\mega\joule\per\kilogram}, and are presented in the same way as in Figure 3 of the main text.
\label{fig:extfig1} }
\end{extendedfigure}

\begin{extendedfigure}
\centering
\includegraphics[width=0.95\linewidth]{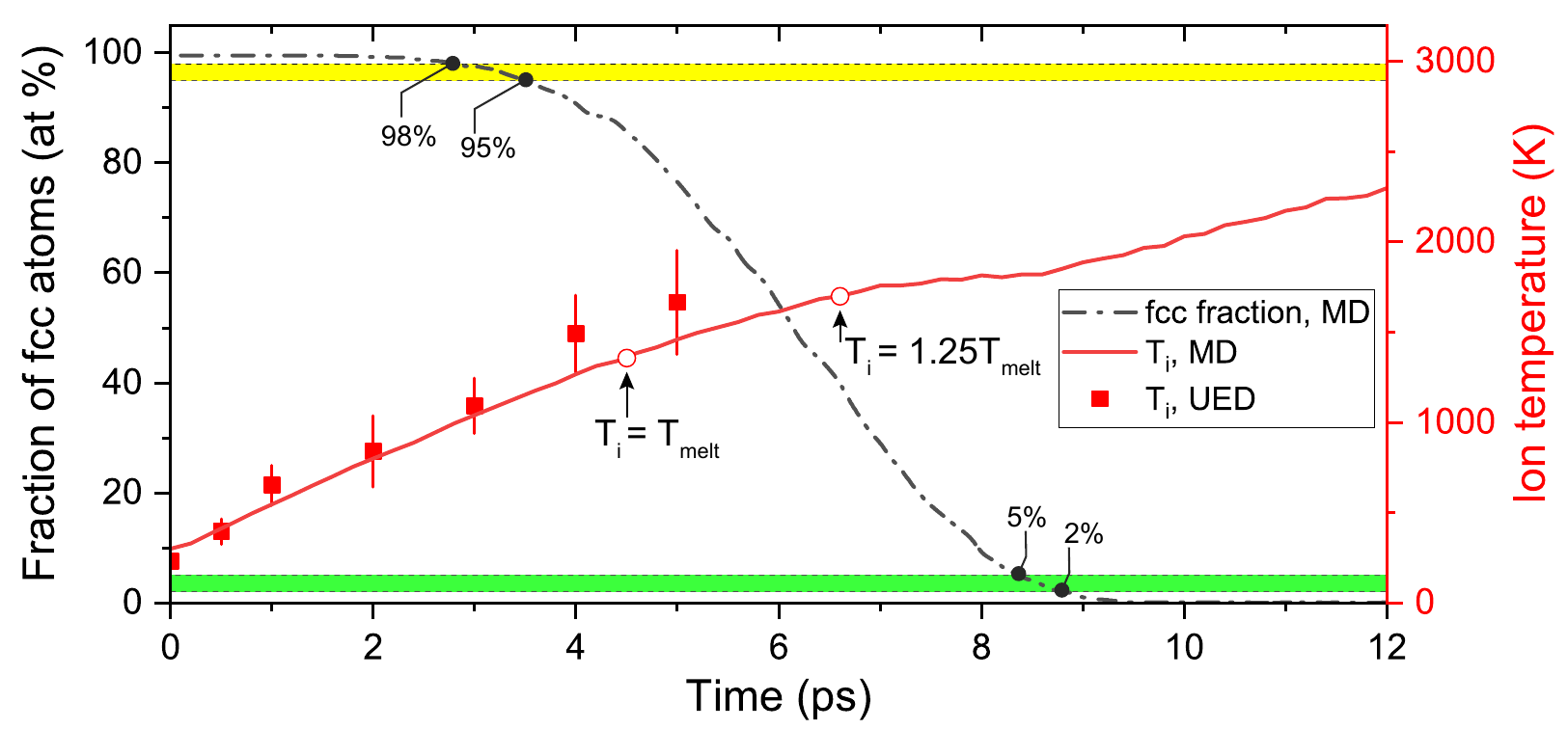}
\caption{\linespread{1.0}\selectfont{}\textbf{TTM-MD simulation results of ultrafast melting in fs laser-heated copper.} The results are presented in the same way as Fig.4c of the main text, but for \(\epsilon\) =  \SI{1.26}{\mega\joule\per\kilogram}.
\label{fig:extfig2} }
\end{extendedfigure}

\end{document}